\documentclass[letterpaper, 10 pt, conference]{IEEEconf}
\IEEEoverridecommandlockouts
\usepackage{cite}
\usepackage{amsmath,amssymb,amsfonts}
\usepackage{algorithmic}
\usepackage{graphicx}
\usepackage{textcomp}
\usepackage{mathtools}
\usepackage{multirow}
\usepackage[table,xcdraw]{xcolor}
\usepackage{adjustbox}
\usepackage{booktabs}
\usepackage{breqn}

\def\BibTeX{{\rm B\kern-.05em{\sc i\kern-.025em b}\kern-.08em
    T\kern-.1667em\lower.7ex\hbox{E}\kern-.125emX}}

\title{\LARGE \bf Unsupervised Deformable Image Registration for Respiratory Motion Compensation in Ultrasound Images

\thanks{$^{1}$Robotics Institute, Carnegie Mellon University, Pittsburgh, PA, USA {\tt\small (abhiman2, aorekhov, choset, jgaleotti)@andrew.cmu.edu} }
}

\author{FNU Abhimanyu$^{1}$, Andrew L. Orekhov$^{1}$, John Galeotti$^{1}$, Howie Choset$^{1}$}

\begin{document}
\maketitle

\begin{abstract} 

In this paper, we present a novel deep-learning model for deformable registration of ultrasound images and an unsupervised approach to training this model. Our network employs recurrent all-pairs field transforms (RAFT) and a spatial transformer network (STN) to generate displacement fields at online rates ($\sim$30\;Hz) and accurately track pixel movement. We call our approach \emph{unsupervised recurrent all-pairs field transforms} (U-RAFT). In this work, we use U-RAFT to track pixels in a sequence of ultrasound images to cancel out respiratory motion in lung ultrasound images. We demonstrate our method on \emph{in-vivo} porcine lung videos. We show a reduction of 76\% in average pixel movement in the porcine dataset using respiratory motion compensation strategy. We believe U-RAFT is a promising tool for compensating different kinds of motions like respiration and heartbeat in ultrasound images of deformable tissue.
\end{abstract}

\section{Introduction} \label{sec:intro}

\par Ultrasound is a widely used modality for medical imaging due to its safety, lack of radiation exposure, and non-invasive nature. With high spatial resolution and real-time imaging capabilities, it can effectively locate malignant tissues for cancer treatment. However, in cases where the tumor is located in the lung or liver, respiratory motion can affect the accuracy of tumor localization. This can result in the exposure of healthy tissues to high levels of radiation, leading to adverse effects. Existing methods like gating and breath-hold can help with tumor localization, but they require significant patient cooperation and setup time. Accurately tracking pixels under respiratory motion and canceling the motion remains a challenging problem.

One approach to respiratory motion compensation is to use deformable registration. Deformable registration is the problem of how to register pairs of images, one referred to as the \emph{fixed image} and the other as the \emph{moving image}, where the two images are of the same anatomy but exhibit displacement between them. For respiratory motion compensation, we can use deformable registration between a fixed image and images during the respiratory cycle to track pixels and compensate for their motion. 

Previous approaches for respiratory motion compensation have tried multiple methods like \cite{renault2005posteriori} used independent component analysis to estimate respiratory kinetics, \cite{averkiou2010quantification} tried to minimize the effects by manually selecting the reference position of the diaphragm, and \cite{mule2011automatic} tried to study the respiratory motion using principal component analysis. However, all these methods treat the respiration process as a rigid process, which causes drift in tracking respiratory motion. In order to cover this limitation, our method treats respiration motion compensations as a non-linear/deformable registration problem.

To compensate for respiratory motion, here we propose a deep learning-based deformable registration approach called Unsupervised-Recurrent All-Pairs Field Transforms (U-RAFT), where we learn to predict the movement of the pixels in an unsupervised manner.
\begin{figure}[ht]
  \centering
        \includegraphics[width=0.75\columnwidth]{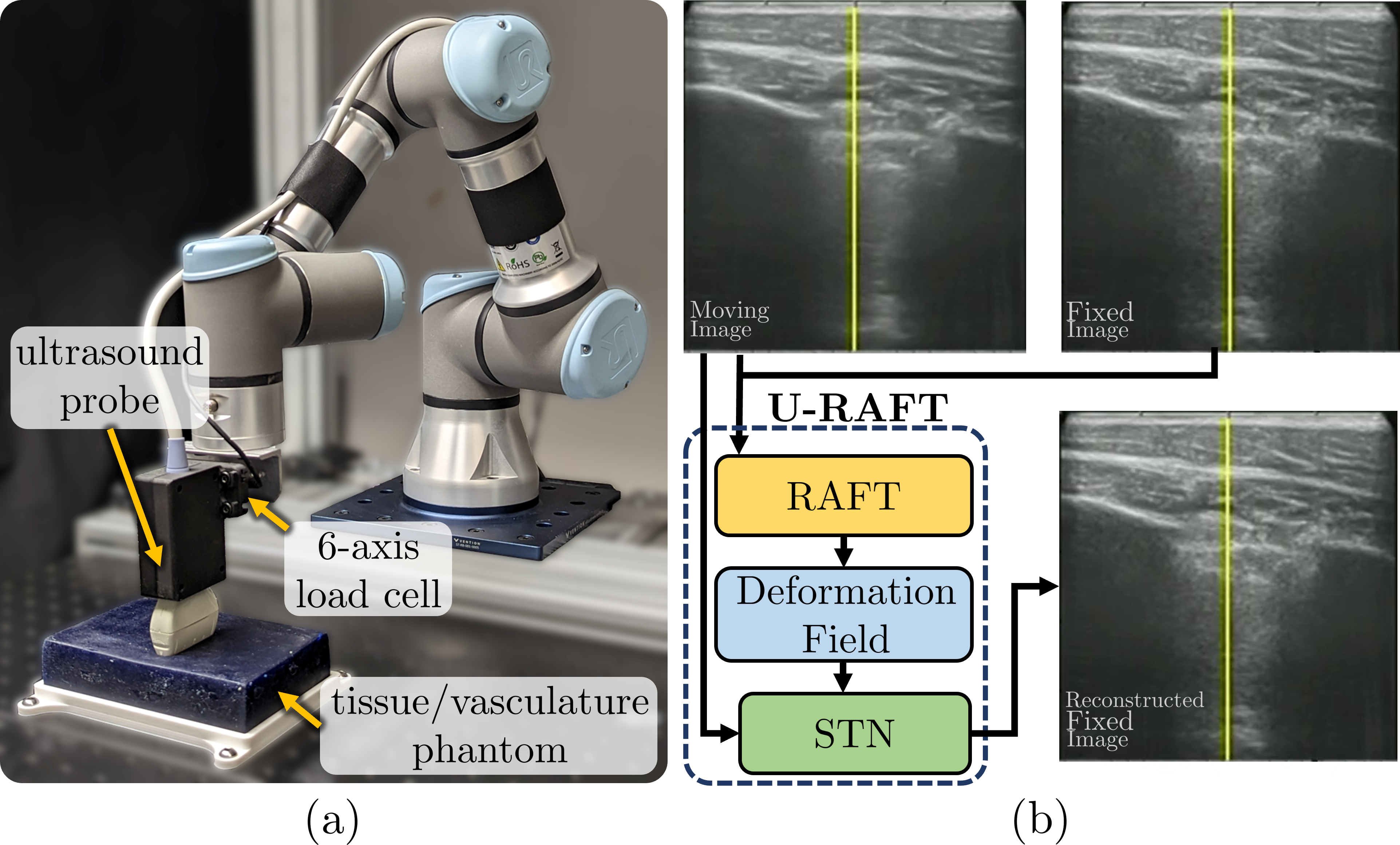}
    \caption{(a) The robot arm we used for capturing the ultrasound images with force-controlled scanning, together with a human tissue and vasculature phantom model. (b) Our U-RAFT model registers a fixed image to a moving image with RAFT, creating a deformation field that is then used with a spatial transformer network (STN) to generate new fixed images, shown here with example \emph{in-vivo} porcine images.}
    \label{fig:system}
\end{figure}

\par In Section \ref{sec:approach}, we present the network architecture of U-RAFT and its use for analyzing and compensating the breathing motion. In Section \ref{sec:results}, we present experimental results using U-RAFT on videos collected on the porcine animals. Section \ref{sec:conclusions} presents our conclusions and discussion on future directions.

\section{Approach}\label{sec:approach}

This section discusses the network architecture and loss function to train the network for predicting the deformation field (DF) and the use of U-RAFT for tracking pixels to capture and compensate for respiratory motion.

\subsection{U-RAFT network architecture and loss function}
Let $I_f$ and $I_m$ be the fixed and moving ultrasound images, respectively from the same respiratory cycle. We denote a displacement field (DF) as $u_{fm} = g_{\theta}(I_f, I_m)$, where $g_{\theta}$ is the function we seek to model with our network and the subscript $\theta$ denotes RAFT's \cite{teed2020raft} network parameters. RAFT is a state-of-the-art optical flow estimation network, but no prior work has shown the application of RAFT on ultrasound images, which are inherently noisier than RGB images\cite{che2017ultrasound}. RAFT is also needs labelled ground truth displacement fields for training. Acquiring ground truth for ultrasound images is time-consuming and labor-intensive, so we seek to make the training unsupervised. We do this by passing the output of RAFT through a Spatial Transformer Network (STN) \cite{NIPS2015_33ceb07b} to generate a reconstructed deformed image $I_f' = \textrm{STN}(u_{fm}, I_{m})$. We then use a cyclic loss function, denoted as ($\mathcal{L}_{\textrm{cyclic-us}}$), to compare $I_f$ to $I_f'$ and $I_m$ to $I_m'$ using multi-scale structural similarity (MS-SSIM), where $I_f'=\textrm{STN}(u_{fm}, I_{m})$ and $I_m'=\textrm{STN}(u_{mf'}, I_{f'})$. We also add a smoothness loss term to regularize the flow prediction.

\subsection{Respiratory compensation using U-RAFT}

We use U-RAFT to track the displacement of every pixel in the video sequence and therefore capture the periodic movement of the tissue during respiratory motion. The first frame in the video sequence is $I_f$ and the rest of the frames are the moving images $I_{m}$. For the frame $i$ in the video sequence, we calculate the net displacement of the pixel at location $x^*,y^*$ as, $d_{x^*,y^*,i} = ||u_{f,m_i}(x^*,y^*)||$, where $||.||$ is the $L_2$ norm. We then pass the sequence of moving images $I_{m_i}$ to STN along with $u_{f,m_i}$ to retrieve the static sequence of images $I_{f_i}' = STN(u_{f,m_i},I_{m_i}) $. Furthermore, we compute the Fast Fourier Transform (FFT) of $d_{x^*,y^*}$ over time to calculate the per-pixel displacement frequency.
\section{Results}\label{sec:results}

\subsection{Experimental setup and training details}
\par The study collected ultrasound images on three live anesthetized pigs for experiments in a controlled lab setting under the supervision of clinicians. The vascular ultrasound images from two of the pigs were used for training a U-RAFT, while lung ultrasound images from the third pig were used for testing. A robotic ultrasound system was used for data collection, including a UR3e manipulator, a Fukuda Denshi portable point-of-care ultrasound scanner, and a six-axis force/torque sensor. The training dataset was collected while palpating with the robot in a sinusoidal force profile, with a minimum and maximum force of 2 N and 10 N, respectively, while the testing dataset was collected by keeping the ultrasound probe static.

\begin{table}[ht]
\begin{center}
\caption{Average pixel displacement before and after applying respiratory compensation}
\label{tab:Video compensation}
\begin{tabular}{|c|c|c|}
\hline
\textbf{Dataset} & \textbf{\begin{tabular}[c]{@{}c@{}}Avg. displacement \\ before compensation (pixel)\end{tabular}} & \textbf{\begin{tabular}[c]{@{}c@{}}Avg. displacement \\ after compensation (pixel)\end{tabular}} \\ \hline
Pig 1 & 2.09 & 0.32 \\ \hline
Pig 2 & 2.32 & 0.65 \\ \hline
Pig 3 & 1.75 & 0.49 \\ \hline
\end{tabular}
\end{center}
\end{table}
\vspace{-0.01in}
\subsection{Analyzing and canceling respiratory motion with U-RAFT}
\begin{figure} [ht]
    \begin{center}
        \includegraphics[width=0.85\linewidth]{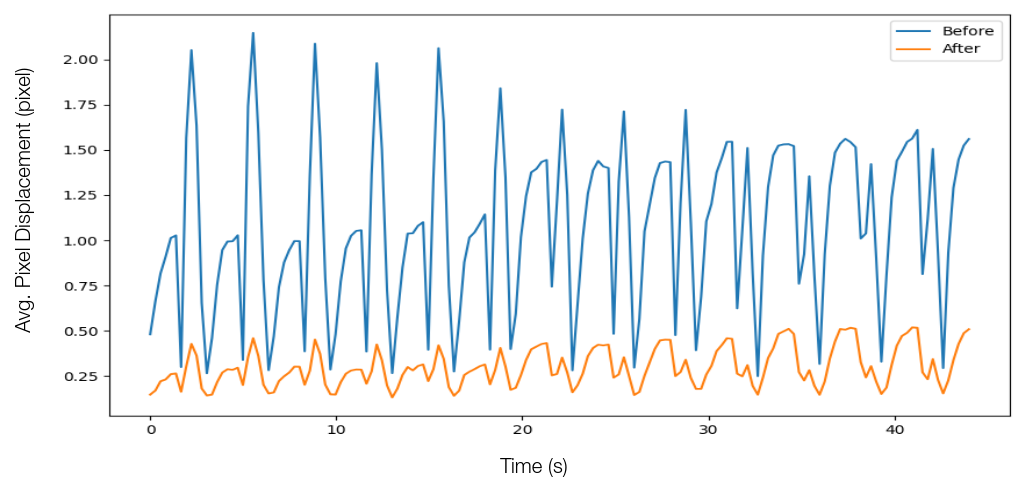}  
    \end{center}
    \caption{Average pixel displacement before and after applying respiratory compensation}. \
    \label{fig:compensation_result}
\end{figure}

Table \ref{tab:Video compensation} provides the mean displacement of pixels for all the frames before and after applying for respiratory compensation. Figure \ref{fig:compensation_result} shows the net displacement of one of the pixels for the whole video sequence around a critical landmark before and after the application of the compensation. Furthermore, we also calculate the respiration rate, via the FFT, by tracking pixels near the rib cage in the ultrasound images. We calculated the respiration rate to be approximately $18$\,beats per minute, which is consistent with the recorded breathing rate of approximately $19$\,bpm during the experiment.

\section{Conclusions and Future Work} \label{sec:conclusions}
In this work we proposed a new deep-learning model for deformable registration of ultrasound images, which we call U-RAFT. The model enables accurate pixel-tracking in at online rates, which makes it suitable for compensating for tissue motion, such as motion due to respiration. The effectiveness of the proposed approach is demonstrated through experiments on multiple \emph{in-vivo} porcine lung videos, where a significant reduction in pixel movement is achieved, and the respiration rate could be estimated from the image alone. These results suggest that the proposed model has potential in helping interpret ultrasound images for roboti-assisted medical inverventions.

\bibliographystyle{IEEEtran}
\bibliography{references}

\end{document}